\documentclass{article}
\usepackage{amsmath}
\usepackage{amssymb}
\usepackage{amsthm}
\usepackage{natbib}
\usepackage{enumitem}
\usepackage{booktabs}

\newcommand{\Z}{\mathbb{Z}}
\newcommand{\F}{\mathbb{F}}
\newcommand{\lep}{\mathrel{\overset{+}{\le}}}
\newcommand{\eqp}{\mathrel{\overset{+}{=}}}
\newcommand{\gep}{\mathrel{\overset{+}{\ge}}}
\newcommand{\pair}[1]{\langle #1 \rangle}

\theoremstyle{plain}
\newtheorem{theorem}{Theorem}
\newtheorem{proposition}{Proposition}
\newtheorem{lemma}{Lemma}
\newtheorem{corollary}{Corollary}
\theoremstyle{definition}
\newtheorem{definition}{Definition}
\newtheorem{example}{Example}
\theoremstyle{remark}
\newtheorem{remark}{Remark}

\title{Individuality from information and intervention:\\ lineages, separation and persistence in cellular automata}
\author{Konstantin Doubrovinski \thanks{correspondence: Konstantin.Doubrovinski@UTSouthwestern.edu}}
\date{\today}

\begin{document}
\maketitle

\begin{abstract}
We propose a notion of individuality for deterministic dynamical systems, developed for cellular automata, based on algorithmic information and interventions. A lineage is a sequence of regions of space-time whose contents determine each other up to a few bits, and into which perturbations of earlier regions are carried forward. Two exact copies of an object cannot be told apart by their contents; they are distinguished by their causal relations, since perturbations of one need not reach the other. Individuals are classes of lineages between which perturbations pass, and they form a genealogy with fission, fusion and reticulation. Chains of cheap steps define a directed bottleneck ultrametric, so the threshold yields a filtration and a persistence structure that grades individuality. No region is prescribed: the descendants of a region are the regions that can be located by a short program, recover its perturbations, are connected, and are of nearly minimal size. Considering all source regions at once gives a canonical nested structure, and an autonomy condition marks preferred scales. All notions are invariant under computable isomorphisms of the intervention structure and use no metric on space. We show that, up to constants, the interventional conditions are pointwise versions of known interventional Shannon quantities, while continuity of lineages is genuinely algorithmic: whether it localizes an object depends on the complexity of that object. We prove that neither ingredient suffices alone: systems with identical runs, and even identical dependence structure, can differ in their individuals. For the parity rule on the square lattice we prove that a parent of side q splits into four separated individuals by time q, that the split persists over a range of thresholds of order $q^2$, and that at large powers of two all descendant regions lie at the four fronts.
\end{abstract}

\section{Introduction}

What makes a part of a dynamical system an individual, and what makes it the same individual at a later time? In biology the question arises for organisms, colonies, symbionts and viruses; in artificial life it arises for patterns in cellular automata, which have no membranes or other built-in boundaries. A recent proposal by Krakauer, Bertschinger, Olbrich, Flack and Ay \cite{krakauer2020} treats individuals as aggregates that propagate information from their past into their future, and formalizes this with Shannon information over ensembles of random variables.

We develop a related notion based on algorithmic information (Kolmogorov complexity) and interventions. It is important to be precise about what is new. The local criteria, transmission and separation of perturbations, quantify over all perturbations of a region, with the rest of the configuration fixed; we show (Proposition \ref{prop:shannon}) that, up to additive constants, they are conditions on the fibres of the response map, i.e. pointwise counterparts of the interventional information flow of Ay and Polani \cite{aypolani2008} for a uniform perturbation and a fixed background (see also \cite{lizier2010}). In this respect the present notion is an algorithmic version of known interventional quantities, and it is pointwise in the background and in the history, but not in the perturbations. Three ingredients have no counterpart in an ensemble formulation. First, \emph{continuity}: lineages are histories along which the actual content changes by few bits per step, a condition on individual strings that depends on the complexity of the actual object (Proposition \ref{prop:pointwise}). Second, \emph{tracking}: the cost of locating a region is charged as the length of a program that finds it, which selects descendant regions intrinsically (Section \ref{sec:descendants}). Third, \emph{relativity to the dynamical law}: all quantities are conditioned on the rule, so that what the rule supplies is not attributed to the individual. On top of these, the notion is \emph{genealogical} and \emph{topological}: it tracks individuals over long times through fission and fusion, and it comes with a filtration, a notion of homotopy for histories, and a persistence structure.

The interventional ingredient addresses a basic difficulty. Two exact copies of an object have the same content, and each determines the other with a constant number of bits; no information-theoretic property of their contents tells them apart. This is the identity of indiscernibles in algorithmic form. Copies are nonetheless distinguished by their \emph{causal relations}: a perturbation of one copy changes its own future and not the future of the other. We use this to define separation between individuals, measured by algorithmic mutual information under interventions. The problem of copies does not arise for ensemble measures of individuality that consider one candidate aggregate at a time, and it is the starting point here.

The topological ingredient comes from a bottleneck construction. Two regions of space-time are linked at level $k$ if they are joined by a chain of steps, each of cost at most $k$. The minimal such $k$ is a directed ultrametric, of the type used to compare programs or proofs by chains of transformations of bounded complexity. Varying $k$ gives a filtration of the genealogy, in which individuals appear, split and merge; the range of $k$ over which a split persists is a graded measure of individuality.

\emph{Contributions.} We define links, lineages, transmission and separation, relate the latter to Shannon quantities, and show that continuity depends on the complexity of the individual object (Sections \ref{sec:lineages}--\ref{sec:separation}); prove that neither information about the run nor the record of which interventions make a difference suffices on its own (Section \ref{sec:necessity}); define descendant windows intrinsically, without a prescribed window shape (Section \ref{sec:descendants}); show that all notions are invariant under computable isomorphisms of the intervention structure, with no metric on space (Section \ref{sec:invariance}); remove the choice of source windows by a canonical nested structure over all of them, with autonomy marking preferred scales (Section \ref{sec:hierarchy}); define individuals and their genealogy, with branching, fusion and a reticulation number (Section \ref{sec:individuals}); introduce homotopy of histories and the persistence structure (Section \ref{sec:topology}); and prove a fission theorem for the parity rule, including the location of all descendant windows at large powers of two (Section \ref{sec:parity}). We illustrate the notions with the ship of Theseus (Section \ref{sec:examples}), state the adequacy conditions the definitions meet and compare with other approaches (Section \ref{sec:adequacy}), and discuss limitations and open problems (Section \ref{sec:discussion}).

\section{Preliminaries}\label{sec:prelim}

\subsection{Algorithmic information}

Strings are finite binary sequences, identified with natural numbers via a fixed computable bijection; $\pair{x,y}$ denotes a computable pairing. We fix a universal prefix-free machine $U$. The (prefix) Kolmogorov complexity of $x$ given $y$ is $K(x\,|\,y) = \min\{ l(p) \,|\, U(p,y) = x\}$, and $K(x) = K(x\,|\,\varepsilon)$ for the empty string $\varepsilon$. We write $a \lep b$ if $a \le b + c$ for a constant $c$ that does not depend on the strings involved (it may depend on $U$, on the chosen encodings, and on explicitly fixed computable functions), and $a \eqp b$ if $a \lep b$ and $b \lep a$. We use the following standard facts \cite{li2008, gacs1974, chaitin1975}.
\begin{enumerate}[label=(F\arabic*)]
\item For every computable $g$, $K(g(x,y)\,|\,y) \lep K(x\,|\,y)$. In particular $K(\pair{x,y}\,|\,z) \lep K(x\,|\,z) + K(y\,|\,z)$, and $K(x\,|\,z) \lep K(y\,|\,z) + K(x\,|\,\pair{y,z})$.
\item (Symmetry of information.) There is a constant $c$ such that $|K(\pair{x,y}\,|\,z) - K(x\,|\,z) - K(y\,|\,\pair{x,z})| \le c\log(n+2)$, where $n = l(x)+l(y)+l(z)$.
\item (Counting.) For every $z$ and $L$, fewer than $2^{L}$ strings $x$ satisfy $K(x\,|\,z) < L$. Hence, for any finite set $X$, more than half of the elements $x \in X$ satisfy $K(x\,|\,z) \ge \log_2 |X| - 1$.
\end{enumerate}
The conditional mutual information is $I(x : y\,|\,z) = K(x\,|\,z) + K(y\,|\,z) - K(\pair{x,y}\,|\,z)$; by (F1), $I(x:y\,|\,z) \lep K(y\,|\,z)$, and by (F2), $I(x:y\,|\,z) = K(x\,|\,z) - K(x\,|\,\pair{y,z})$ up to $O(\log(n+2))$. The conditional information distance \cite{bennett1998} is
\[
E(x,y\,|\,z) = \max\{ K(x\,|\,\pair{y,z}),\, K(y\,|\,\pair{x,z})\} \quad .
\]
It satisfies $E(x,y\,|\,z) \lep E(x,w\,|\,z) + E(w,y\,|\,z)$ by (F1).

\subsection{Cellular automata, light cones and interventions}

Fix $d \ge 1$, a finite alphabet $\Sigma$ with a quiescent state $0$, a radius $\rho \ge 1$ and a local rule $\phi : \Sigma^{\{-\rho,\ldots,\rho\}^d} \to \Sigma$ with $\phi(0,\ldots,0) = 0$. The global map is $F(c)(i) = \phi((c(i+e))_{|e|_\infty \le \rho})$ for configurations $c : \Z^d \to \Sigma$. A configuration is finite if its support $\operatorname{supp}(c) = \{i \,|\, c(i) \ne 0\}$ is finite; $F$ maps finite configurations to finite configurations. The run of a finite configuration $c_0$ is $c_t = F^t(c_0)$, $t \ge 0$. Rules, finite configurations, and restrictions of configurations to finite sets (as arrays with their position) are encoded by strings in a fixed computable way.

No metric on $\Z^d$ is used in the definitions below; all geometry is derived from the dynamics. The \emph{dependence set} of a cell $w$ is
\[
N(w) = \{ u \in \Z^d \,|\, F(c)(w) \ne F(c')(w) \text{ for some } c, c' \text{ that differ only at } u\} \quad .
\]
It is contained in $w + \{-\rho,\ldots,\rho\}^d$ and computable from $\phi$ and $w$. Since a function on a finite product does not depend on a coordinate if changing that coordinate alone never changes its value, $F(c)(w)$ depends only on $c|_{N(w)}$. For $R \subseteq \Z^d$, let $R^{\leftarrow} = \bigcup_{w \in R} N(w)$ (the cells on which $R$ depends) and $R^{\rightarrow} = \{ w \,|\, N(w) \cap R \ne \emptyset\}$ (the cells that $R$ influences), and let $R^{\leftarrow t}$ be the $t$-fold iterate of $R \mapsto R^{\leftarrow}$, with $R^{\leftarrow 0} = R$.

\begin{lemma}[Light cone]\label{lem:cone}
If $c$ and $c'$ agree on $R^{\leftarrow t}$, then $F^t(c)$ and $F^t(c')$ agree on $R$.
\end{lemma}
\begin{proof}
By induction on $t$: if $c$, $c'$ agree on $R^{\leftarrow (t+1)} = (R^{\leftarrow})^{\leftarrow t}$, then $F^t(c)$, $F^t(c')$ agree on $R^{\leftarrow}$, and $F(\cdot)(i)$ for $i \in R$ depends only on values in $N(i) \subseteq R^{\leftarrow}$.
\end{proof}

\emph{Cellular systems.} The definitions of Sections \ref{sec:lineages}--\ref{sec:topology} use only the following structure: a set of cells, a finite alphabet with a quiescent state, a global map $F$ on finite configurations, and the dependence sets $N(w)$, which are finite, computable from $w$, with $F(c)(w)$ depending only on $c|_{N(w)}$ and each cell in only finitely many, computably given, sets $N(w)$. We call such a structure a \emph{cellular system}, and its cells with their alphabets its \emph{intervention structure}: the interventions considered are the replacements of the contents of finite sets of cells. Translation invariance and the lattice structure of $\Z^d$ are used only in the examples. The relevant notion of equivalence, under which all definitions are invariant, is that of Section \ref{sec:invariance}.

An \emph{intervention} at time $s$ replaces the configuration $c_s$ of a run by a modified configuration $c'_s$ and continues the run from there: $c'_{s+\tau} = F^\tau(c'_s)$. In a cellular automaton every intervention in the sense of Pearl's do-operator \cite{pearl2009} is of this form.

\section{Sites, links and lineages}\label{sec:lineages}

Regions of space-time are described by finite windows of arbitrary shape. No bound on their size is fixed in advance. Without some constraint the whole configuration would trivially be one individual; the constraints are supplied by information and interventions (Definition \ref{def:transmission} and Section \ref{sec:descendants}), not by a prescribed window shape.

\begin{definition}[Sites]\label{def:site}
A site is a pair $\sigma = (s, W_\sigma)$ with $s \ge 0$ a time and $W_\sigma \subseteq \Z^d$ a finite nonempty set, its window. Its content in the run $(c_t)$ is the array $v_\sigma = c_s|_{W_\sigma}$; as all restrictions, it is encoded together with its positions, so $W_\sigma$ is computable from $v_\sigma$. For sites $\sigma = (s,W)$, $\tau = (s',W')$, the frame is $\theta(\sigma,\tau) = \pair{\phi, s, s'}$: the rule and the two times.
\end{definition}

\begin{definition}[Links and chains]\label{def:link}
A step is a pair of sites $\sigma = (s,W)$, $\sigma' = (s+1, W')$ with $W' \subseteq W^{\rightarrow}$ and $W \subseteq W'^{\leftarrow}$: every cell of the new window depends on a cell of the old one, and every cell of the old window influences a cell of the new one. Its cost is the information distance between the contents, given the frame:
\[
\ell(\sigma, \sigma') = E\bigl(v_\sigma, v_{\sigma'} \,\big|\, \theta(\sigma,\sigma')\bigr) \quad .
\]
A chain from $\sigma$ to $\tau$ is a sequence of sites $\sigma = \sigma_0, \sigma_1, \ldots, \sigma_n = \tau$ in which each consecutive pair is a step; its cost is $\max_i \ell(\sigma_i, \sigma_{i+1})$ (and $0$ if $n = 0$). The linking cost from $\sigma$ to a later site $\tau$ is
\[
D(\sigma, \tau) = \min \{ \text{cost of } \gamma \,|\, \gamma \text{ a chain from } \sigma \text{ to } \tau \} \quad ,
\]
with $D(\sigma,\tau) = \infty$ if no chain exists.
\end{definition}

The cost of a step measures how much of the new content is not determined by the old one, and how much of the old content is lost. Since contents carry their positions, the cost includes the description of each window from the other; for a translation of a window to a neighbouring position this is a constant. The frame deliberately does not contain the windows: otherwise the choice of a window's shape, among the many shapes compatible with a step, could encode information about the content, and the cost would not measure what it is meant to measure. A chain of cost at most $k$ is a history along which content changes by at most $k$ bits per step; over many steps, it may change completely.

\begin{proposition}[Directed ultrametric]\label{prop:ultra}
For sites $\sigma$, $\pi$, $\tau$ with times $s(\sigma) \le s(\pi) \le s(\tau)$:
\begin{enumerate}[label=(\roman*)]
\item $D(\sigma,\sigma) = 0$;
\item $D(\sigma,\tau) \le \max\{ D(\sigma,\pi),\, D(\pi,\tau)\}$.
\end{enumerate}
For every $k$, the relation ``$D(\sigma,\tau) \le k$'' is therefore reflexive and transitive: a preorder on space-time, respecting the direction of time.
\end{proposition}
\begin{proof}
(i) The chain of length $0$. (ii) If $\gamma_1$ is a chain from $\sigma$ to $\pi$ and $\gamma_2$ a chain from $\pi$ to $\tau$, their concatenation is a chain from $\sigma$ to $\tau$ whose cost is the maximum of their costs.
\end{proof}

\begin{remark}[Bottleneck distances]
$D$ is a bottleneck (minimax) distance: the cost of a connection is that of its most expensive step, minimized over all connections. Its symmetric counterpart on a set of objects connected by transformations of bounded complexity, where the distance is the least $k$ such that two objects are joined by a chain of transformations each of complexity at most $k$, is an ultrametric, and its sublevel sets are the clusters of single-linkage clustering; their evolution with $k$ is $0$-dimensional persistent homology \cite{carlsson2009, edelsbrunner2010}. Here the transformations are single time steps of the dynamics, and the resulting structure is directed.
\end{remark}

In a closed system, links are cheap wherever the information of the initial object is locally present. Hence continuity alone constrains where information resides, not how it is processed; the substantive conditions are those of Section \ref{sec:separation}.

\begin{lemma}[Cheap links in closed runs]\label{lem:closed}
Suppose $c_0$ vanishes outside a finite set $A$, and let $x = c_0|_A$ (with its position). There is a constant $c$ such that, for every step $(\sigma,\sigma')$ with windows $W$, $W'$, $K(x \,|\, \pair{\theta, v_\sigma}) \le k$ and $K(x \,|\, \pair{\theta, v_{\sigma'}}) \le k$, where $\theta = \theta(\sigma,\sigma')$, we have $\ell(\sigma,\sigma') \le k + \max\{K(W'\,|\,W), K(W\,|\,W')\} + c$.
\end{lemma}
\begin{proof}
The run is computable from $x$ and $\theta$ (which contains the rule and the times), so $v_{\sigma'}$ is computable from $x$, $\theta$ and $W'$. By (F1), $K(v_{\sigma'}\,|\,\pair{\theta, v_\sigma}) \lep K(x\,|\,\pair{\theta,v_\sigma}) + K(W'\,|\,W)$, since $W$ is computable from $v_\sigma$; and symmetrically.
\end{proof}

Information about the content can be lost along a chain only gradually:

\begin{lemma}[Information along chains]\label{lem:chain}
There is a constant $c$ such that for every chain $\sigma_0, \ldots, \sigma_j$ of cost at most $k$, with $\sigma_0$ at time $s_0$,
\[
K\bigl(v_{\sigma_0} \,\big|\, \pair{\phi, s_0, j, v_{\sigma_j}}\bigr) \le j\,\bigl(k + 2\log_2(j+1) + c\bigr) \quad .
\]
\end{lemma}
\begin{proof}
For each $i < j$, $K(v_{\sigma_i}\,|\,\pair{\theta_i, v_{\sigma_{i+1}}}) \le k$, where $\theta_i = \pair{\phi, s_0 + i, s_0 + i + 1}$ is computable from $\pair{\phi,s_0}$ and $i$, and $K(i) \le 2\log_2(j+1) + c$. Composing the $j$ programs by (F1) gives the bound.
\end{proof}

The following proposition shows that continuity is a condition on the individual object, not on a distribution of objects. For a finite set $A$ and $t \ge 0$, let $A^{\rightarrow t}$ be the $t$-fold iterate of $R \mapsto R^{\rightarrow}$; by Lemma \ref{lem:cone} and $\phi(0,\ldots,0) = 0$, a run that vanishes outside $A$ at time $0$ vanishes outside $A^{\rightarrow t}$ at time $t$.

\begin{proposition}[Continuity depends on the individual parent]\label{prop:pointwise}
There is a constant $c$ such that the following holds. Suppose $c_0$ vanishes outside a finite set $A$, let $x = c_0|_A$ (with its position), $\beta = \log_2|\Sigma|$, and $\sigma_0 = (0, A)$.
\begin{enumerate}[label=(\roman*)]
\item Every step $(\sigma, \sigma')$, with windows $W$, $W'$, has cost $\ell(\sigma,\sigma') \le K(x\,|\,\phi) + \max\{K(W'\,|\,W), K(W\,|\,W')\} + c$. In particular, if $K(x\,|\,\phi) \le k - c - w$, then every family of sites in which consecutive windows form steps and determine each other with at most $w$ bits, e.g. translates of a fixed window moving to neighbouring positions, is a chain of cost at most $k$, wherever it lies, including regions outside $A^{\rightarrow t}$ in which the run is quiescent.
\item If $K(x\,|\,\phi) \ge |A|\beta - d$, and $\gamma$ is a chain of cost at most $k$ from $\sigma_0$ to a site $\tau$ at time $j$ whose window lies outside $A^{\rightarrow j}$, then
\[
j\,\bigl(k + 2\log_2(j+1) + c\bigr) \ \ge\ |A|\beta - d - K(W_\tau\,|\,\pair{\phi, j}) - 2\log_2(j+1) - c \quad .
\]
\end{enumerate}
\end{proposition}
\begin{proof}
(i) Every content is computable from $x$, the rule, the time and the window; argue as in Lemma \ref{lem:closed}. (ii) The content of $\tau$ is the quiescent array on $W_\tau$, so $K(\pair{j, v_\tau}\,|\,\phi) \le K(j) + K(W_\tau\,|\,\pair{\phi,j}) + c$. By (F1), $K(x\,|\,\phi) \le K(x\,|\,\pair{\phi, 0, j, v_\tau}) + K(\pair{j,v_\tau}\,|\,\phi) + c$, and Lemma \ref{lem:chain} bounds the first term, since $v_{\sigma_0} = x$.
\end{proof}

For a simple parent, continuity does not localize anything: cheap chains run everywhere, including through empty space, and only the interventional conditions of Section \ref{sec:separation} tell a lineage from a drifting empty window. For an incompressible parent, a cheap chain cannot leave the region where the parent's information resides faster than about $k$ bits per step; it needs at least about $|A|\beta/k$ steps to reach empty space. Both statements concern a single object $x$. In an ensemble formulation with a random parent, the first case has vanishing weight, and averaged quantities cannot express that the localizing power of continuity depends on which parent is present.

\section{Transmission and separation}\label{sec:separation}

For a site $\sigma = (s,W_\sigma)$ and an array $\delta \in \Sigma^{W_\sigma}$, let $c^{\sigma\leftarrow\delta}$ be the run obtained by the intervention that replaces the content of $W_\sigma$ at time $s$ by $\delta$, leaving $c_s$ unchanged outside $W_\sigma$. For a site $\tau = (s',W_\tau)$ with $s' \ge s$, let $v^\delta_\tau = c^{\sigma\leftarrow\delta}_{s'}|_{W_\tau}$ be the response, and let
\[
z = z(\sigma,s') = \pair{\phi,\, s,\, W_\sigma,\, s',\, c_s|_{\Z^d \setminus W_\sigma}}
\]
be the \emph{source data}: the rule, the source site, the target time, and the unperturbed configuration outside the source window at time $s$ (a finite configuration). The source data do not contain the target window.

\begin{definition}[Transmission and separation]\label{def:transmission}
Let $k \ge 0$, and let $\tau$ be a site not earlier than $\sigma$.
\begin{enumerate}[label=(\roman*)]
\item $\sigma$ \emph{transmits} to $\tau$ at level $k$, written $\sigma \Rightarrow_k \tau$, if $K(W_\tau\,|\,z) \le k$ and, for at least $3/4$ of all $\delta \in \Sigma^{W_\sigma}$, $K(\delta \,|\, \pair{z, v^\delta_\tau}) \le k$: the target window can be located from the source data with at most $k$ bits (\emph{tracking}), and the response there determines the perturbation (\emph{sufficiency}).
\item $\sigma$ is \emph{separated} from $\tau$ at level $k$, written $\sigma \perp_k \tau$, if for at least $3/4$ of all $\delta \in \Sigma^{W_\sigma}$, $I(\delta : v^\delta_\tau \,|\, z) \le k$: the response at $\tau$ carries almost no information about the perturbation.
\end{enumerate}
\end{definition}

Both notions combine an intervention with an information measurement, and both are upper bounds on information. They refer to most perturbations, so they can be established by counting, without proving that any particular string is complex. Both are monotone: $\sigma \Rightarrow_k \tau$ implies $\sigma \Rightarrow_{k'} \tau$, and $\sigma \perp_k \tau$ implies $\sigma \perp_{k'} \tau$, for $k \le k'$. The tracking condition is what replaces a prescribed window shape: it requires that the descendant be found by a short program from what is known about the source, and it excludes windows assembled cell by cell from wherever the information happens to be (Remark \ref{rem:why}). For windows carrying more than about $2k$ bits the two notions are mutually exclusive:

\begin{lemma}[Transmission and separation are opposite]\label{lem:opposite}
There is a constant $c$ such that if $\sigma \Rightarrow_k \tau$ and $\sigma \perp_k \tau$, then $|W_\sigma| \log_2 |\Sigma| \le 2k + c \log(n+2) + 1$, where $n$ bounds the lengths of $z$, $\delta$ and $v^\delta_\tau$.
\end{lemma}
\begin{proof}
By (F3), more than half of the $\delta \in \Sigma^{W_\sigma}$ satisfy $K(\delta\,|\,z) \ge |W_\sigma|\log_2|\Sigma| - 1$. Since each of the two conditions holds for at least $3/4$ of all $\delta$, some $\delta$ satisfies all three properties. For it, by (F2), $I(\delta : v^\delta_\tau\,|\,z) \ge K(\delta\,|\,z) - K(\delta\,|\,\pair{z,v^\delta_\tau}) - c\log(n+2) \ge |W_\sigma|\log_2|\Sigma| - 1 - k - c\log(n+2)$, which must be at most $k$.
\end{proof}

Transmission and separation have close Shannon counterparts. Fix $\sigma$, $\tau$ and the source data $z$, and let $g : \Sigma^{W_\sigma} \to \Sigma^{W_\tau}$, $g(\delta) = v^\delta_\tau$, be the response map; it is computable from $z$ and $W_\tau$. Let $N = |\Sigma|^{|W_\sigma|}$, and for a response $v$ let $g^{-1}(v)$ be its fibre. If $\delta$ is uniformly distributed on $\Sigma^{W_\sigma}$ and $V = g(\delta)$, then $H(\delta\,|\,V) = \mathbb{E}\log_2|g^{-1}(V)|$ and $I(\delta;V) = H(V) = \mathbb{E}\log_2(N/|g^{-1}(V)|)$; these are the interventional information quantities of \cite{aypolani2008} for a uniform intervention on $W_\sigma$ with the background held fixed.

\begin{proposition}[Shannon counterparts]\label{prop:shannon}
There is a constant $c$ such that, for all $\sigma$, $\tau$, $z$ as above, with $n$ bounding the lengths of $z$, $\delta$ and $v^\delta_\tau$:
\begin{enumerate}[label=(\roman*)]
\item for every $\delta$, $K(\delta\,|\,\pair{z, v^\delta_\tau}) \le \log_2|g^{-1}(v^\delta_\tau)| + c$;
\item for every $d \ge 0$, fewer than a fraction $2^{-d}$ of all $\delta$ satisfy $K(\delta\,|\,\pair{z,v^\delta_\tau}) < \log_2|g^{-1}(v^\delta_\tau)| - d$;
\item $H(\delta\,|\,V) \le \mathbb{E}\,K(\delta\,|\,\pair{z,V}) \le H(\delta\,|\,V) + c$, for $\delta$ uniform;
\item for every $\delta$, $I(\delta : v^\delta_\tau\,|\,z) \le \log_2(N/|g^{-1}(v^\delta_\tau)|) + K(W_\tau\,|\,z) + c$; and for every $d \ge 0$, all but a fraction $2^{-d}$ of all $\delta$ satisfy $I(\delta : v^\delta_\tau\,|\,z) \ge \log_2(N/|g^{-1}(v^\delta_\tau)|) - d - c\log(n+2)$.
\end{enumerate}
\end{proposition}
\begin{proof}
(i) From $z$ and $v = v^\delta_\tau$, which contains $W_\tau$, the fibre $g^{-1}(v)$ is computable, and $\delta$ is specified by its index in the fibre, a string of known length $\lceil\log_2|g^{-1}(v)|\rceil$. (ii) For each response $v$, by (F3) fewer than $2^{-d}|g^{-1}(v)|$ elements of the fibre have complexity, given $\pair{z,v}$, below $\log_2|g^{-1}(v)| - d$; sum over $v$. (iii) The upper bound is the average of (i). For the lower bound, for each $v$ the lengths $K(\delta\,|\,\pair{z,v})$, $\delta \in g^{-1}(v)$, satisfy Kraft's inequality, so their average over the fibre is at least $\log_2|g^{-1}(v)|$; average over $v$. (iv) By (F1) and (F2), $I(\delta:v\,|\,z) \lep K(v\,|\,z)$, and $v$ is described from $z$ by $W_\tau$ and the Shannon--Fano code of $v$ for the distribution of $V$, which is computable from $z$ and $W_\tau$ and has length $\log_2(N/|g^{-1}(v)|) + O(1)$. For the lower bound, by (F2) $I(\delta:v\,|\,z) \ge K(\delta\,|\,z) - K(\delta\,|\,\pair{z,v}) - c\log(n+2)$; by (F3), all but a fraction $2^{-d}$ of $\delta$ have $K(\delta\,|\,z) \ge \log_2 N - d$, and (i) bounds the second term.
\end{proof}

Thus, given the source data, the sufficiency part of transmission at level $k$ holds when at least $3/4$ of all perturbations lie in fibres of size at most $2^{k-c}$, and fails when more than $1/4 + 2^{-d}$ of them lie in fibres of size greater than $2^{k+d}$; separation is the corresponding statement for large fibres. Up to these constants, transmission and separation are conditions on the response map alone, and the Kolmogorov complexity in them can be replaced by counting. What does not reduce in this way is the tracking condition, which charges for locating the target window, and the continuity of Definition \ref{def:link}, which concerns the actual contents rather than perturbations (Proposition \ref{prop:pointwise}).

\begin{lemma}[Size, support and number of targets]\label{lem:size}
There is a constant $c$ such that the following holds for every site $\sigma = (s, W_\sigma)$, time $s' \ge s$ and $k \ge 0$. Write $\beta = \log_2 |\Sigma|$, and let $R = R(\sigma,s')$ be the set of cells $i$ such that $c^{\sigma\leftarrow\delta}_{s'}(i)$ depends on $\delta \in \Sigma^{W_\sigma}$.
\begin{enumerate}[label=(\roman*)]
\item If $\sigma \Rightarrow_k (s', W)$, then $|W|\, \beta \ge |W_\sigma|\, \beta - 2k - c$.
\item If $\sigma \Rightarrow_k (s', W)$, then $|W \cap R|\, \beta \ge |W_\sigma|\, \beta - 3k - c$.
\item Fewer than $2^{k+1}$ windows $W$ satisfy $\sigma \Rightarrow_k (s', W)$.
\end{enumerate}
\end{lemma}
\begin{proof}
Let $z = z(\sigma, s')$. (iii) Every such $W$ satisfies $K(W\,|\,z) \le k$, and by (F3) fewer than $2^{k+1}$ strings do.

(i) By (F3) and the definition of transmission, some $\delta$ satisfies both $K(\delta\,|\,z) \ge |W_\sigma|\beta - 1$ and $K(\delta\,|\,\pair{z,v^\delta}) \le k$, where $v^\delta$ is the response on $W$. The response consists of $W$ and an element of $\Sigma^{W}$, so by (F1) $K(v^\delta\,|\,z) \lep K(W\,|\,z) + |W|\beta \le k + |W|\beta$. By (F1), $K(\delta\,|\,z) \lep K(v^\delta\,|\,z) + K(\delta\,|\,\pair{z,v^\delta})$, and the claim follows.

(ii) The set $R$ is computable from $z$, by running the intervention for each of the finitely many $\delta$; and the values of the response outside $R$ do not depend on $\delta$, so they are computable from $z$ and $W$ (run the intervention for any fixed $\delta$). Let $u^\delta$ be the response restricted to $W \cap R$. Then $v^\delta$ is computable from $z$, $W$ and $u^\delta$, so $K(\delta\,|\,\pair{z,u^\delta}) \lep K(W\,|\,z) + K(\delta\,|\,\pair{z,v^\delta}) \le 2k$ for at least $3/4$ of all $\delta$. Since $W \cap R$ is computable from $z$ and $W$, $K(u^\delta\,|\,z) \lep k + |W\cap R|\beta$. Now argue as in (i).
\end{proof}

\begin{remark}[Why tracking and economy are needed]\label{rem:why}
Transmission is roughly monotone in the target window: if $\sigma \Rightarrow_k (s',W)$, $W \subseteq W'$ and $K(W'\,|\,z) \le k$, then the sufficiency condition holds for $W'$ at level $2k + c$, since $v^\delta_{W}$ is computed from $v^\delta_{W'}$ and $W$, and $K(W\,|\,\pair{z, W'}) \lep K(W\,|\,z)$. Large simple windows, such as a ball of radius $2^n$ (complexity about $\log n$) or the whole light cone, therefore transmit as soon as some small window inside them does; descendants must be required to be of nearly minimal size. Conversely, without the tracking condition, a set of cells chosen one by one from all over the light cone, each carrying some of the information of the perturbation, would count as a descendant; under a linear rule such as that of Section \ref{sec:parity}, such scattered sets exist in abundance, and locating one costs about $\log_2\binom{N}{|W|}$ bits, where $N$ is the number of cells in the light cone.
\end{remark}

\section{Neither information nor intervention suffices}\label{sec:necessity}

The definitions combine an information measure with interventions. We now show that neither ingredient can be dropped: a criterion that sees only the run, with any information measure applied to it, cannot tell interacting copies from non-interacting ones; and a criterion that sees the run together with the full record of which interventions affect which regions, but not how much information they carry, cannot tell a lineage from a channel that destroys its content. Both constructions are one-dimensional, with radius $1$.

We call \emph{dependence structure} of a run the set of all pairs $((s,U),(s',W'))$, with $s \le s'$ and $U$, $W'$ finite, such that the response on $W'$ at time $s'$ to an intervention on $U$ at time $s$ is a nonconstant function of the perturbation $\delta \in \Sigma^U$. It records which interventions make a difference where, and nothing about how much.

\begin{proposition}[Necessity of both ingredients]\label{prop:necessity}
There is a constant $c$ such that the following holds.
\begin{enumerate}[label=(\roman*)]
\item \emph{(Information alone does not suffice.)} For every $q \ge 1$, $L \ge q$ and $x \in \F_2^q$ there are two cellular automata $F_1$, $F_2$ with alphabet $\Sigma = \F_2^2$ and an initial configuration $c_0$ such that $F_1^t(c_0) = F_2^t(c_0)$ for all $t \ge 0$; the run contains, at all times, exact copies of $x$ in the windows $A = \{0,\ldots,q-1\}$ and $B = \{L, \ldots, L+q-1\}$; under $F_1$, $(0,A) \perp_c (t,B)$ for every $t$, the response on $B$ not depending on the perturbation at all; and under $F_2$, $(0,A) \not\perp_k (L,B)$ for every $k < q - c\log(L+q+2)$.
\item \emph{(Intervention alone does not suffice.)} For every $m \ge 2$, $q \ge 1$ and $x \in \{0, e_1\}^q \subseteq (\F_2^m)^q$ there are two cellular automata $G_1$, $G_2$ with alphabet $\Sigma = \F_2^m$ and an initial configuration $c_0$ such that the two runs coincide, the two runs have the same dependence sets and the same dependence structure, and the family $\Lambda(s) = (s, W + s)$, $W = \{0,\ldots,q-1\}$, carries the same contents in both; under $G_1$, $\Lambda$ is a $k$-lineage on every interval for every $k \ge c$; and under $G_2$, $\Lambda(s) \not\Rightarrow_k \Lambda(s')$ for all $s < s'$ and all $k < (m-1)q - c$, so that $\Lambda$ is not a $k$-lineage on any interval of positive length.
\end{enumerate}
\end{proposition}
\begin{proof}
(i) Write states as pairs $(a,b) \in \F_2^2$. Let $F_1$ be the identity, and let $F_2$ keep the first component and shift the second to the right: $F_2(c)(i) = (a(i), b(i-1))$. Both have quiescent state $(0,0)$. Let $c_0$ have first components $x$ on $A$ and on $B$ and $0$ elsewhere, and second components $0$ everywhere. Since $F_2$ shifts only the second components, which vanish, both runs are constant, and they coincide. Under $F_1$, an intervention on $A$ changes nothing outside $A$, so the response on $B$ is the unperturbed content, computable from $z$, and $I(\delta : v^\delta\,|\,z) \le c$ by (F1). Under $F_2$, write $\delta = (\delta_a, \delta_b)$. The second components of $\delta$ travel to the right, one cell per step, so at time $L$ the response on $B$ consists of the first components of $x$ and the translate of $\delta_b$. Hence $K(\delta\,|\,\pair{z,v^\delta}) \le K(\delta_a) + c \le q + 2\log_2 q + c$, while by (F3) at least $3/4$ of all $\delta$ satisfy $K(\delta\,|\,z) \ge 2q - 2$. By (F2), these $\delta$ satisfy $I(\delta:v^\delta\,|\,z) \ge q - c\log(L+q+2)$ for a suitable constant $c$, so separation fails at every level below this bound.

(ii) Let $G_1(c)(i) = c(i-1)$, and $G_2(c)(i) = P(c(i-1))$ with $P(y) = (y_1 + \cdots + y_m)\, e_1$. Both have quiescent state $0$, and both have dependence sets $N(w) = \{w - 1\}$, since $P$ is nonconstant in each coordinate. Let $c_0 = x$ on $W$ and $0$ elsewhere. Since $P$ fixes $0$ and $e_1$, both runs are $c_t(i) = x(i - t)$, so they coincide, and $\Lambda$ carries the translates of $x$. For an intervention on $U$ at time $s$, the response at time $s' = s + t$ is $c_{s'}(w) = \delta(w - t)$ under $G_1$ and $c_{s'}(w) = P^t(\delta(w-t))$ under $G_2$ for $w \in U + t$, and the unperturbed value elsewhere. For $t \ge 1$, $P^t = P$ is nonconstant on $\F_2^m$, so in both systems the response on $W'$ is nonconstant if and only if $W' \cap (U + t) \neq \emptyset$; for $t = 0$ both responses are $\delta$ on $U$. The dependence structures therefore coincide. Under $G_1$, consecutive sites of $\Lambda$ form steps of constant cost, and the response on $W + s'$ is the translate of $\delta$, so $\Lambda(s) \Rightarrow_c \Lambda(s')$; hence $\Lambda$ is a $k$-lineage for $k \ge c$. Under $G_2$, for $s' > s$, the response on $W + s'$ is the array of parities $(\sum_j \delta(i)_j)\, e_1$, $i \in W + s$; each fibre of the response map has $2^{(m-1)q}$ elements. By Proposition \ref{prop:shannon}(ii) with $d = 2$, more than $1/4$ of all $\delta$ satisfy $K(\delta\,|\,\pair{z,v^\delta}) \ge (m-1)q - 2$, so transmission fails for $k < (m-1)q - 2$.
\end{proof}

Part (i) concerns any criterion that is a function of the run alone: whatever information measure it applies to the contents, it assigns the same verdict to the two systems, although in one the copies cannot affect each other and in the other a perturbation of the first copy reaches the second with about $q$ bits. Part (ii) concerns any criterion that is a function of the run and its dependence structure: it assigns the same verdict to a system in which the moving pattern carries its perturbations forward intact and to one in which they are compressed to one bit per cell. The criterion of Definition \ref{def:transmission} distinguishes both pairs because it measures, for interventions, how much information arrives. Proposition \ref{prop:necessity} is thus a minimality statement for the two ingredients, not a uniqueness statement for the particular definitions; it parallels the corresponding necessity results for self-reproduction in companion work.

\section{Descendant windows}\label{sec:descendants}

\begin{definition}[Connected windows]\label{def:connected}
The \emph{influence graph} has the cells as vertices, with $u$ and $w$ adjacent if $u \in N(w)$ or $w \in N(u)$. A window $W$ is \emph{connected} if the subgraph of the influence graph induced on $W$ is connected.
\end{definition}

Connectedness is an interventional condition. If $W = A \sqcup B$ with no edge between $A$ and $B$, then over one time step no intervention on $A$ changes the next state of any cell of $B$, and conversely: the parts do not interact directly. Conversely, if there is an edge, say $u \in A$ with $u \in N(w)$, $w \in B$, then some intervention at $u$ changes the next state of $w$ in some configuration. Connectedness is therefore exactly the absence of a split into parts that cannot affect each other in one step. Proposition \ref{prop:chimera} below shows that it cannot be dropped.

\begin{definition}[Descendant windows]\label{def:descendant}
Let $k, \lambda \ge 0$, let $\sigma$ be a site at time $s$, and let $s' \ge s$. Let
\[
m_k(\sigma, s') = \min\{ |W| \,|\, W \text{ connected},\ \sigma \Rightarrow_k (s', W)\} \quad ,
\]
with $m_k = \infty$ if no such $W$ exists. A \emph{$(k,\lambda)$-descendant window} of $\sigma$ at time $s'$ is a connected window $W$ with $\sigma \Rightarrow_k (s', W)$ and $(|W| - m_k(\sigma,s'))\log_2|\Sigma| \le \lambda$.
\end{definition}

A descendant window thus satisfies four conditions: it can be located from the source data with few bits (tracking), it determines the perturbations of the source (sufficiency), it is connected, and it is within $\lambda$ bits of the smallest size of connected windows with the first two properties (economy). The economy condition removes surplus cells and also connected windows that join pieces of distinct copies by long paths (Theorem \ref{thm:intrinsic}).

\begin{proposition}[Basic properties]\label{prop:basic}
There is a constant $c$ such that, for every site $\sigma$ and time $s' \ge s$: there are fewer than $2^{k+1}$ $(k,\lambda)$-descendant windows of $\sigma$ at time $s'$; each has at least $|W_\sigma| - (2k+c)/\log_2|\Sigma|$ cells, of which at least $|W_\sigma| - (3k+c)/\log_2|\Sigma|$ lie in the set $R(\sigma,s')$ of cells affected by perturbations of $\sigma$; and if some connected $W$ satisfies $\sigma \Rightarrow_k (s',W)$ and $|W|\log_2|\Sigma| \le |W_\sigma|\log_2|\Sigma| + \lambda - 2k - c$, then $W$ is a $(k,\lambda)$-descendant window.
\end{proposition}
\begin{proof}
The first two statements are Lemma \ref{lem:size}. For the last, Lemma \ref{lem:size}(i) gives $m_k(\sigma,s')\log_2|\Sigma| \ge |W_\sigma|\log_2|\Sigma| - 2k - c$, so $(|W| - m_k)\log_2|\Sigma| \le \lambda$.
\end{proof}

All conditions in Definition \ref{def:descendant} are relative to the source window $W_\sigma$, which specifies what is perturbed and is still chosen. What is no longer chosen is where, and in what shape, its descendants are sought. In Section \ref{sec:hierarchy} the choice of source window is removed altogether, by considering all source windows at once.

\section{Invariance}\label{sec:invariance}

\begin{definition}[Computable isomorphisms]\label{def:iso}
Let $(V, \Sigma, F)$ and $(V', \Sigma', G)$ be cellular systems. A \emph{computable isomorphism} between them is a pair $(\pi, h)$, where $\pi : V \to V'$ is a bijection with $\pi$ and $\pi^{-1}$ computable, and $h$ assigns to each cell $i \in V$ a bijection $h_i : \Sigma \to \Sigma'$ with $h_i(0) = 0$, computably in $i$, such that the induced map on configurations, $H(c)(\pi(i)) = h_i(c(i))$, satisfies $G \circ H = H \circ F$ on finite configurations.
\end{definition}

A computable isomorphism may relabel cells in an arbitrary computable way, as long as it maps the interventions of one system, replacements of the contents of sets of cells, onto those of the other. It extends to sites by $H(s, W) = (s, \pi(W))$, and to lineages sitewise.

\begin{proposition}[Invariance]\label{prop:invariance}
Let $(\pi,h)$ be a computable isomorphism, with induced map $H$. There is a constant $c_H$, depending only on $H$, the two rules and the encodings, such that for every run, all sites $\sigma$, $\tau$ and all $k$, $\lambda$:
\begin{enumerate}[label=(\roman*)]
\item $N_G(\pi(w)) = \pi(N_F(w))$ for every cell $w$; hence $\pi$ preserves the influence graph, connectedness, steps and the sets of affected cells, and $|\pi(W)| = |W|$;
\item $|\ell(H\sigma, H\sigma') - \ell(\sigma,\sigma')| \le c_H$ for every step $(\sigma,\sigma')$;
\item $\sigma \Rightarrow_k \tau$ implies $H\sigma \Rightarrow_{k + c_H} H\tau$, and $\sigma \perp_k \tau$ implies $H\sigma \perp_{k+c_H} H\tau$; the same holds for $H^{-1}$;
\item every $k$-lineage is mapped to a $(k + c_H)$-lineage, and $\Lambda \approx_s \Lambda'$ at level $k$ implies $H\Lambda \approx_s H\Lambda'$ at level $k + c_H$;
\item if $W$ is a $(k,\lambda)$-descendant window of $\sigma$ at time $s'$, then $\pi(W)$ is a $(k + c_H, \lambda')$-descendant window of $H\sigma$ at time $s'$, with $\lambda' = \lambda + (m_k(\sigma,s') - m_{k+2c_H}(\sigma,s'))\log_2|\Sigma|$.
\end{enumerate}
\end{proposition}
\begin{proof}
(i) Configurations $c$, $c'$ differ only at $u$ if and only if $H(c)$, $H(c')$ differ only at $\pi(u)$, and $G(H(c))(\pi(w)) = h_w(F(c)(w))$ with $h_w$ a bijection. The remaining statements follow, since all of these notions are defined from the dependence sets and the cardinalities of windows. (Note that $|\Sigma'| = |\Sigma|$.)

(ii)--(iii) The frame, the source data, the contents, the perturbations and the responses of the image are obtained from those of the original by computable bijections, uniformly given by $H$ and the codes of the rules: the source data $z'$ of $H\sigma$ are computable from $z$ and conversely, the target window $\pi(W)$ from $W$ and conversely, and the perturbation $\delta \mapsto (h_i(\delta(i)))_i$ is a bijection, so fractions of perturbations are preserved. By (F1), every conditional complexity in Definitions \ref{def:link} and \ref{def:transmission} changes by at most a constant, and so does every mutual information, which is a combination of three such terms.

(iv) follows from (i)--(iii) and the definitions.

(v) Connectedness and transmission are preserved by (i) and (iii). By (iii) applied to $H^{-1}$, every connected $W'$ with $H\sigma \Rightarrow_{k+c_H} (s', W')$ satisfies $\sigma \Rightarrow_{k + 2c_H} (s', \pi^{-1}(W'))$, so $m_{k+c_H}(H\sigma, s') \ge m_{k + 2c_H}(\sigma, s')$. Hence $(|\pi(W)| - m_{k+c_H}(H\sigma,s'))\log_2|\Sigma| \le (|W| - m_k(\sigma,s'))\log_2|\Sigma| + (m_k(\sigma,s') - m_{k+2c_H}(\sigma,s'))\log_2|\Sigma|$.
\end{proof}

Thus all notions are invariant up to constant shifts of the thresholds, except that economy is invariant only where the minimal size $m_k$, which is nonincreasing in $k$, does not drop between $k$ and $k + 2c_H$. In the language of persistence, the two filtrations are interleaved: the structure at level $k$ of one system is contained in that at level $k + c_H$ of the other, and conversely. Features that persist over an interval of thresholds much longer than $c_H$, such as the fission of Corollary \ref{cor:persistence}, are therefore invariant.

Computable isomorphisms do not include local recodings that change the cells themselves, such as sliding-block codes or higher-block presentations, which merge several cells into one and introduce forbidden blocks; nor arbitrary computable bijections of whole configurations, which do not preserve the intervention structure. No notion of the present kind can be invariant under the latter: a single run, considered as a sequence of configurations up to arbitrary computable bijections, carries no decomposition into parts, so that two copies cannot be told from one (Section \ref{sec:discussion}).

\section{Individuals and genealogy}\label{sec:individuals}

\begin{definition}[Lineages]\label{def:lineage}
A $k$-lineage on a time interval $[s_0, s_1]$ is a family of sites $\Lambda = (\Lambda(s))_{s_0 \le s \le s_1}$ with $\Lambda(s)$ at time $s$, such that
\begin{enumerate}[label=(L\arabic*)]
\item \emph{(continuity)} $(\Lambda(s), \Lambda(s+1))$ is a step of cost at most $k$, for $s_0 \le s < s_1$;
\item \emph{(integrity)} $\Lambda(s) \Rightarrow_k \Lambda(s')$ for all $s_0 \le s \le s' \le s_1$.
\end{enumerate}
\end{definition}

Continuity says that the content changes slowly; integrity says that the lineage carries its own variations forward. A lineage is thus a bounded region of space-time through which information propagates, and in which perturbations persist.

To compare lineages, we group them, at each time, according to whether perturbations can pass between them.

\begin{definition}[Individuals and genealogy]\label{def:genealogy}
Let $\mathcal{L}$ be a finite family of $k$-lineages on a common interval $[s_0,s_1]$. For $s \in [s_0,s_1]$ and $\Lambda, \Lambda' \in \mathcal{L}$, write $\Lambda \approx_s \Lambda'$ if $\Lambda(s) = \Lambda'(s)$, or $\Lambda(s) \Rightarrow_k \Lambda'(s')$ for some $s' \in [s, s_1]$, or $\Lambda'(s) \Rightarrow_k \Lambda(s')$ for some $s' \in [s,s_1]$. The equivalence classes of the transitive closure of $\approx_s$ are the \emph{$k$-individuals of $\mathcal{L}$ at time $s$}. The \emph{genealogy} $G_k(\mathcal{L})$ is the graph whose vertices are the pairs $(s, C)$, with $C$ an individual at time $s$, and whose edges join $(s, C)$ and $(s+1, C')$ whenever $C \cap C' \neq \emptyset$.
\end{definition}

An individual at time $s$ is thus a maximal group of lineages between which perturbations can pass, directly or through intermediaries. A vertex of the genealogy with two successors is a \emph{fission} (division or budding); a vertex with two predecessors is a \emph{fusion}. Distinct lineages may also belong to the same individual, e.g. two parts of one organism.

\begin{definition}[Reticulation]
The reticulation number of $\mathcal{L}$ at level $k$ is the cycle rank $b_1(G_k(\mathcal{L})) = |E| - |V| + \kappa$, where $E$, $V$ and $\kappa$ are the numbers of edges, vertices and connected components of $G_k(\mathcal{L})$.
\end{definition}

The family $\mathcal{L}$ in Definition \ref{def:genealogy} can be chosen intrinsically, as the family of all lineages that consist of descendant windows of a given site.

\begin{definition}[Descendant lineages]\label{def:desclineage}
Let $\sigma_0$ be a site at time $s_0$, and $s_1 \ge s_0$. A \emph{$(k,\lambda)$-descendant lineage} of $\sigma_0$ on $[s_0,s_1]$ is a $k$-lineage $\Lambda$ with $\Lambda(s_0) = \sigma_0$ such that, for every $s \in (s_0, s_1]$, the window of $\Lambda(s)$ is a $(k,\lambda)$-descendant window of $\sigma_0$ at time $s$. Their set is denoted $\mathcal{D}_{k,\lambda}(\sigma_0, s_1)$.
\end{definition}

By Proposition \ref{prop:basic}, $\mathcal{D}_{k,\lambda}(\sigma_0,s_1)$ is finite, with fewer than $2^{(k+1)(s_1 - s_0)}$ elements, so $G_k(\mathcal{D}_{k,\lambda}(\sigma_0,s_1))$ is defined. It depends only on $\sigma_0$, $s_1$, $k$ and $\lambda$: the \emph{intrinsic genealogy} of $\sigma_0$.

A genealogy without fusions is a forest, and has reticulation number $0$. Cycles arise when lineages that have separated later exchange information, as in recombination or horizontal transfer, so the reticulation number measures how far the genealogy departs from a tree.

\begin{remark}[Copies and the identity of indiscernibles]
Two lineages that carry exact copies of the same object have contents that determine each other at constant cost, so no condition on contents can tell them apart. Definition \ref{def:genealogy} tells them apart by the propagation of perturbations, a relational and causal property. By Lemma \ref{lem:opposite}, separation and transmission exclude each other for windows carrying more than about $2k$ bits, so two such lineages that are mutually separated at all times from some point on are distinct individuals.
\end{remark}

\section{Homotopy and persistence}\label{sec:topology}

\begin{definition}[Homotopy of chains]\label{def:homotopy}
Two chains $\gamma = (\sigma_0, \ldots, \sigma_n)$ and $\gamma' = (\sigma'_0, \ldots, \sigma'_n)$ with the same endpoints ($\sigma_0 = \sigma'_0$, $\sigma_n = \sigma'_n$) differ by an \emph{elementary deformation at level $k$} if they differ in exactly one intermediate site, and both have cost at most $k$. They are \emph{homotopic at level $k$} if they are joined by a finite sequence of elementary deformations at level $k$.
\end{definition}

Homotopy at level $k$ is an equivalence relation on chains of cost at most $k$ with fixed endpoints, and it is monotone in $k$: chains homotopic at level $k$ are homotopic at every level $k' \ge k$. Inequivalent chains between the same endpoints are distinct routes by which information travels from one site to another, separated by regions in which no cheap chain exists. Chains respect the direction of time, so the natural framework is that of directed homotopy, developed for spaces with an irreversible time direction, e.g. in concurrency theory \cite{grandis2009, fajstrup2006}.

\emph{Persistence.} All the notions of Sections \ref{sec:lineages}--\ref{sec:individuals} depend on a threshold $k$. Continuity and integrity become easier to satisfy as $k$ grows, so more families of sites qualify as lineages; separation also becomes easier, so more pairs of lineages count as separated. These effects work in opposite directions on the number of individuals. A clean way to organize them is to use two thresholds, $k$ for continuity and integrity and $k'$ for separation, giving a two-parameter family of genealogies $G_{k,k'}$. A feature of the genealogy, such as a fission into a given number of individuals, that is present over a large region of the $(k,k')$-plane is robust; one that is present only in a small region is marginal. We call the extent of this region the \emph{persistence} of the feature, and regard it as a graded degree of individuality. In the single-threshold setting of Definition \ref{def:genealogy}, persistence is an interval of values of $k$ (Corollary \ref{cor:persistence}).

\section{Fission under the parity rule}\label{sec:parity}

Let $d = 2$, $\Sigma = \F_2$, and
\[
F(c)(i) = \sum_{e \in N_0} c(i + e) \pmod 2 \quad , \qquad N_0 = \{\pm e_1, \pm e_2\} \quad ,
\]
the parity rule on the four nearest neighbours (usually attributed to Fredkin), with $\rho = 1$. Its dependence sets are $N(w) = w + N_0$, so its influence graph is the $4$-neighbour graph of $\Z^2$; in particular, a connected window is a $4$-connected set of cells. Since $N_0 = -N_0$, consecutive front windows $Q + se$ and $Q + (s+1)e$ form a step. Fix $q \ge 1$, the parent's window $Q = \{0, \ldots, q-1\}^2$, and an arbitrary parent $x \in \F_2^Q$; the run starts from $c_0$ equal to $x$ on $Q$ and $0$ elsewhere. The parent's site is $\sigma_0 = (0, Q)$. For $e \in N_0$ and $s \ge 0$, let the \emph{front site} be $\sigma_e(s) = (s, Q + s e)$, and let $\Lambda_e = (\sigma_e(s))_{s \ge 0}$. All four front lineages start at $\sigma_0$. The box $Q$ only describes the parent; the descendant windows are characterized intrinsically in Theorem \ref{thm:intrinsic}.

Let $N_s(v) \in \F_2$ be the parity of the number of walks of length $s$ with steps in $N_0$ from $0$ to $v \in \Z^2$. Since $F$ is linear over $\F_2$ and $F = \sum_{e\in N_0} T_e$ with $(T_e c)(i) = c(i+e)$, we have $F^s = \sum_v N_s(v) T_v$, i.e.
\begin{equation}\label{eq:kernel}
c_s(i) = \sum_{j \in \Z^2} N_s(j - i)\, c_0(j) \quad .
\end{equation}
Since every walk of length $s$ ends at $\ell_1$-distance at most $s$ from its start, $N_s(v) = 0$ whenever $|v|_1 > s$. Consequently:

\begin{lemma}[$\ell_1$ light cone]\label{lem:l1cone}
For every finite configuration $c$, $\operatorname{supp}(F^\tau(c)) \subseteq \{ j + v \,|\, j \in \operatorname{supp}(c), \, |v|_1 \le \tau\}$.
\end{lemma}

\begin{lemma}[Fronts determine the parent]\label{lem:front}
For every $s \ge 0$ and $e \in N_0$, the map $\F_2^Q \to \F_2^Q$, $y \mapsto F^s(y)|_{Q + se}$ (where $y$ is placed on $Q$ and $0$ elsewhere, and the result is translated back to $Q$) is a linear bijection. Moreover, for every $\Delta \in \F_2^{Q + b}$ placed on $Q + b$, the map $\Delta \mapsto F^\tau(\Delta)|_{Q + b + \tau e}$ is a linear bijection, for every $b \in \Z^2$ and $\tau \ge 0$.
\end{lemma}

\begin{proof}
By the symmetries of the lattice, which commute with $F$, it suffices to take $e = e_1$, and by translation invariance it suffices to prove the first statement. For $i = (s + a, b) \in Q + s e_1$ and $j \in Q$, equation (\ref{eq:kernel}) gives the coefficient $N_s(v)$ with $v = j - i = (j_1 - a - s,\, j_2 - b)$. If $j_1 < a$, then $|v|_1 \ge |v_1| > s$ and the coefficient is $0$. If $j_1 = a$, then $v_1 = -s$, so every step of the walk is $-e_1$, hence $v_2 = 0$, and the coefficient is $1$ if $j_2 = b$ and $0$ otherwise. Ordering both $Q + s e_1$ and $Q$ by their first coordinate (relative to the window), the matrix of the map is therefore block upper triangular with identity blocks on the diagonal, and hence invertible over $\F_2$.
\end{proof}

The proof shows more: for $0 \le a_0 < q$, the cells of $Q + se_1$ with relative first coordinate at least $a_0$ depend only on the cells of $Q$ with first coordinate at least $a_0$, through a block triangular map with identity diagonal blocks. Hence the content of the outer columns $\{a_0, \ldots, q-1\} \times \{0,\ldots,q-1\}$ of the front window determines the corresponding columns of the parent; by symmetry, the same holds for the outer columns or rows of every front window.

\begin{theorem}[Fission into four individuals]\label{thm:fission}
There are constants $c$ and $c'$, depending only on $U$ and the encodings, such that the following holds for every $q \ge 1$ and every parent $x$.
\begin{enumerate}[label=(\roman*)]
\item \emph{(Continuity)} Every step $(\sigma_e(s), \sigma_e(s+1))$ has cost at most $c$.
\item \emph{(Integrity)} For all $s \le s'$ and every $\delta \in \F_2^{Q+se}$, $K(Q + s'e\,|\,z) \le c$ and $K\bigl(\delta \,|\, \pair{z, v^\delta_{\sigma_e(s')}}\bigr) \le c$, where $z = z(\sigma_e(s), s')$ and the intervention is at $\sigma_e(s)$. In particular, each $\Lambda_e$ is a $k$-lineage on every interval, for every $k \ge c$.
\item \emph{(Separation)} For $e \neq e'$, $s \ge q$, $s' \ge s$ and every $\delta \in \F_2^{Q+se}$, the response $v^\delta_{\sigma_{e'}(s')}$ to an intervention at $\sigma_e(s)$ does not depend on $\delta$, and $I(\delta : v^\delta_{\sigma_{e'}(s')} \,|\, z) \le c$, where $z = z(\sigma_e(s), s')$.
\item \emph{(Copies)} For every $t = 2^m \ge q$, the content of each front window $Q + te$ at time $t$ is an exact translate of $x$.
\end{enumerate}
\end{theorem}

\begin{proof}
(i) By Lemma \ref{lem:front}, $v_{\sigma_e(s)} = L_s(x)$ and $v_{\sigma_e(s+1)} = L_{s+1}(x)$ for linear bijections $L_s$, $L_{s+1}$ that are computable from the rule, the times, $q$ and $e$. Each window is obtained from the other by a translation by $\pm e$, and $q$ is read off the window, so each content is computed from the other by a fixed program, applied to the frame and the two bits that specify $e$; the step cost is bounded by a constant.

(ii) Let $W = Q + se$ and $\tau = s' - s$. The target window $Q + s'e = W + \tau e$ is computed from $z$ (which contains $W$, $s$ and $s'$) and the two bits that specify $e$. By linearity, $c^{\sigma\leftarrow\delta}_{s'} = F^\tau(\tilde c_s) + F^\tau(\delta)$, where $\tilde c_s$ is $c_s$ with $W$ set to $0$ and $\delta$ is placed on $W$; the first term is computable from $z$. By Lemma \ref{lem:front}, $F^\tau(\delta)|_{W + \tau e}$ is a linear bijection of $\delta$, computable from the frame. Since $W + \tau e = Q + s'e$, $\delta$ is computed from $z$ and $v^\delta_{\sigma_e(s')}$ by a fixed program, which reads $e$ off the position of the response.

(iii) By Lemma \ref{lem:l1cone}, the perturbation affects only cells at $\ell_1$-distance at most $\tau = s' - s$ from $W = Q + se$. We show that every cell of $Q + s'e'$ is at $\ell_1$-distance greater than $\tau$ from $W$. Let $p, p' \in Q$, so that $|p_l - p'_l| \le q - 1$ in each coordinate. If $e' = -e$, the difference of $se + p$ and $s'e' + p'$ in the coordinate of $e$ has absolute value at least $s + s' - (q-1)$, so the $\ell_1$-distance is at least $2s + \tau - (q-1) > \tau$, since $s \ge q$. If $e' \perp e$, the two coordinates of the difference have absolute values at least $s - (q-1)$ and $s' - (q-1)$, so the $\ell_1$-distance is at least $2s + \tau - 2(q-1) > \tau$. Hence $v^\delta_{\sigma_{e'}(s')} = F^\tau(\tilde c_s)|_{Q+s'e'}$, which does not depend on $\delta$ and is computable from $z$. By (F1), $I(\delta : v^\delta \,|\, z) \lep K(v^\delta\,|\,z) \lep 0$.

(iv) Since $F^2 = \sum_{e,e'} T_{e+e'} = \sum_e T_{2e}$ over $\F_2$ (the terms with $e \neq e'$ cancel in pairs), induction gives $F^{2^m} = \sum_{e \in N_0} T_{2^m e}$. With $t = 2^m$, $c_t(i) = \sum_{e} x(i + te)$. For $i = j + te'$ with $j \in Q$, the term $e = -e'$ gives $x(j)$; for $e \neq -e'$, the point $i + te = j + t(e + e')$ lies outside $Q$, because $e + e' \neq 0$ has a coordinate of absolute value at least $1$ and $t \ge q$. Hence $c_t|_{Q + te'}$ is the translate of $x$.
\end{proof}

\begin{corollary}[Genealogy and persistence]\label{cor:persistence}
Let $\mathcal{L} = \{\Lambda_e \,|\, e \in N_0\}$ on an interval $[0, s_1]$ with $s_1 \ge q$. For every $k$ with $c \le k < q^2 - c'$, the genealogy $G_k(\mathcal{L})$ has exactly one vertex at time $0$ and exactly four vertices at every time $s \in [q, s_1]$. Thus the parent undergoes a fission into four individuals no later than time $q$, and this fission persists over an interval of thresholds of length about $q^2$, the information content of the parent's window.
\end{corollary}

\begin{proof}
By Theorem \ref{thm:fission}(i)--(ii), each $\Lambda_e$ is a $k$-lineage for $k \ge c$. At time $0$ all four lineages are at the site $(0,0)$, so they form one individual. Let $s \ge q$ and $e \neq e'$. By Theorem \ref{thm:fission}(iii), for every $s' \ge s$ the response at $\sigma_{e'}(s')$ to an intervention at $\sigma_e(s)$ is computable from $z$, so $K(\delta\,|\,\pair{z, v^\delta}) \gep K(\delta\,|\,z)$. By (F3), more than half of the $\delta$ satisfy $K(\delta\,|\,z) \ge q^2 - 1$, so if $k < q^2 - c'$ for a suitable constant $c'$, the transmission condition fails for more than a quarter of the $\delta$, and $\sigma_e(s) \not\Rightarrow_k \sigma_{e'}(s')$. Hence $\Lambda_e \not\approx_s \Lambda_{e'}$ for all $e \neq e'$, and the four lineages are four distinct individuals at time $s$.
\end{proof}

We now show that the four fronts are also singled out intrinsically, and that connectedness is needed for this.

\begin{proposition}[Chimeras]\label{prop:chimera}
There is a constant $c$ such that for every $q \ge 2$ and $s \ge 0$ the following holds. Let $h = \lfloor q/2 \rfloor$, and let $C_s$ be the union of the outer $h$ columns $\{s + q - h, \ldots, s + q - 1\} \times \{0,\ldots,q-1\}$ of the east front window and the outer $q - h$ columns $\{-s, \ldots, -s + q - h - 1\} \times \{0,\ldots,q-1\}$ of the west front window. Then $|C_s| = q^2$, $\sigma_0 \Rightarrow_c (s, C_s)$, and for $s \ge 1$ the window $C_s$ is not connected: its two parts lie at $\ell_1$-distance $2s + 1 > 1$. For $s \ge q$ they lie inside the two separated front windows of Theorem \ref{thm:fission}(iii).
\end{proposition}
\begin{proof}
$C_s$ is computable from $q$ and $s$, so $K(C_s\,|\,z) \le c$ for $z = z(\sigma_0,s)$. By the remark after Lemma \ref{lem:front}, the east part determines the columns $q-h, \ldots, q-1$ of the perturbation. Reflection in the line $i_1 = (q-1)/2$ maps $Q$ to itself, commutes with $F$, and exchanges the east and west front windows, mapping the outer $q-h$ columns of the west window onto the outer $q-h$ columns of the east window; hence the west part determines the columns $0, \ldots, q-h-1$. Together they determine the perturbation by a fixed program, so $K(\delta\,|\,\pair{z, v^\delta}) \le c$ for every $\delta$. The distance statement is immediate.
\end{proof}

A chimera has the size of the parent, is located at constant cost, and determines the parent exactly; by size and information content alone it is as good a descendant as a front window. It is assembled from parts of two copies that are separated from each other. Its exclusion is the role of connectedness in Definition \ref{def:descendant}; at large powers of two, the economy condition excludes chimeras whose parts are joined by a path.

\begin{theorem}[Descendant windows of the parity rule]\label{thm:intrinsic}
There is a constant $c$ such that the following holds for every $q \ge 1$, every parent $x$, every $k \ge c$ and every $\lambda \ge 2k + c$.
\begin{enumerate}[label=(\roman*)]
\item For every $s \ge 0$ and $e \in N_0$, the front window $Q + se$ is a $(k,\lambda)$-descendant window of $\sigma_0$ at time $s$. In particular, each front lineage $\Lambda_e$ is a $(k,\lambda)$-descendant lineage of $\sigma_0$ on every interval $[0, s_1]$.
\item Suppose $3k + c < q^2$, and let $t = 2^m$ with $t > q + \lambda + 3k + c$. Then every $(k,\lambda)$-descendant window of $\sigma_0$ at time $t$ meets exactly one of the four front windows $Q + te$, and has at least $q^2 - 3k - c$ cells in it.
\item Suppose moreover $t > q^2 + q + \lambda + 3k + c$. If $\Lambda \in \mathcal{D}_{k,\lambda}(\sigma_0, 2t)$, then the windows of $\Lambda(t)$ and $\Lambda(2t)$ meet the front windows of the same front $e$. Consequently, every $(k,\lambda)$-descendant lineage of $\sigma_0$ meets, at all powers of two $t' \ge t$ in its interval, the front windows of one and the same front.
\end{enumerate}
\end{theorem}
\begin{proof}
Write $z_s = z(\sigma_0, s)$; the background outside $Q$ is $0$, and the response to $\delta \in \F_2^Q$ is $F^s(\delta)$ with $\delta$ placed on $Q$.

(i) The window $Q + se$ is connected, and by Theorem \ref{thm:fission}(ii) (with the intervention at time $0$), $\sigma_0 \Rightarrow_c (s, Q + se)$. By Lemma \ref{lem:size}(i), $m_k(\sigma_0, s) \ge q^2 - 2k - c$, so $q^2 - m_k \le 2k + c \le \lambda$. The step conditions hold since consecutive front windows differ by a translation by one cell, and the lineage conditions hold by Theorem \ref{thm:fission}(i)--(ii).

(ii) By Theorem \ref{thm:fission}(iv), $F^t(\delta) = \sum_{e} T_{te}\delta$, so the set $R(\sigma_0, t)$ of affected cells is contained in the union $B$ of the four front windows. Let $W$ be a $(k,\lambda)$-descendant window. By Lemma \ref{lem:size}(ii), $|W \cap B| \ge q^2 - 3k - c > 0$, so $W$ meets some front window. Suppose $W$ meets two. Since $W$ is connected, it contains a path of cells, each at $\ell_1$-distance $1$ from the next, from one front window to another, and we may choose it so that its intermediate cells lie outside $B$. Distinct front windows are at $\ell_1$-distance at least $2(t - q + 1)$ (attained by perpendicular fronts), so the path has at least $2(t-q) + 1 > t - q$ intermediate cells, and $|W| \ge q^2 - 3k - c + t - q$. By (i), $m_k(\sigma_0, t) \le q^2$, so economy gives $|W| \le q^2 + \lambda$. Hence $t - q \le \lambda + 3k + c$, a contradiction.

(iii) Let $W_t$ and $W_{2t}$ be the windows of $\Lambda(t)$ and $\Lambda(2t)$, and suppose by (ii) that they meet the windows of fronts $e$ and $e'$. By Definition \ref{def:link}, every cell of the window at time $s+1$ depends on a cell of the window at time $s$, hence lies at $\ell_1$-distance $1$ from it; so every cell of $W_{2t}$ lies within $\ell_1$-distance, and hence $\ell_\infty$-distance, $t$ of a cell of $W_t$. If $e' \ne e$, the windows $Q + te$ and $Q + 2te'$ are at $\ell_\infty$-distance at least $2t - q + 1$; hence $W_t$ contains a cell within distance $t$ of $Q + 2te'$ and a cell in $Q + te$, which are at distance at least $t - q + 1$. But $W_t$ is $4$-connected with at most $q^2 + \lambda$ cells, so its $\ell_1$-diameter, and hence its $\ell_\infty$-diameter, is less than $q^2 + \lambda$, a contradiction. The last statement follows by induction over successive powers of two.
\end{proof}

Thus, with the parent's window as the only prescribed region, the descendants at large powers of two are exactly localized at the four fronts, and every descendant lineage follows a single front. What Theorem \ref{thm:intrinsic} leaves open is the intrinsic genealogy itself, i.e. the count of individuals in $G_k(\mathcal{D}_{k,\lambda}(\sigma_0,s_1))$: a lineage that follows front $e$ need not coincide with $\Lambda_e$ between powers of two, and separation between lineages on different fronts requires an analysis of their windows at intermediate times. We conjecture that the intrinsic genealogy has four individuals at all times $s \ge q$ in the range of thresholds of Corollary \ref{cor:persistence}.

\begin{remark}[Numerical observations]
For $q = 4$ we computed, by Gaussian elimination over $\F_2$ applied to the linear map $x \mapsto c_s|_W$, which windows $W$ of shape $Q$ determine the parent. Such windows exist at every time we tested ($0 \le s \le 16$ and $s \in \{24, 31, 32, 33, 48, 63, 64\}$), as Lemma \ref{lem:front} guarantees. We also counted all such windows at the times $s \in \{1, 2, 3, 5, 7, 8, 12, 15, 16, 20, 31, 32\}$. At $s = 1$ there are four, forming one cluster of adjacent positions. At the other times except $12$ and $20$ there are exactly four, one at each front, in four separate clusters. At $s = 12 = 8 + 4$ and $s = 20 = 16 + 4$ there are sixteen; there the configuration is the sum of four translates of the configuration at time $4$ (since $F^{8} = \sum_e T_{8e}$ and $F^{16} = \sum_e T_{16 e}$), each contributing its own four fronts. Thus, at the times tested, no window of the parent's size other than these carries the parent's information. Dropping the shape constraint, we enumerated, for $q = 2$ ($19$ shapes; $s \le 20$) and $q = 3$ ($9910$ shapes; $3 \le s \le 7$), all connected windows (Definition \ref{def:connected}; for this rule, the fixed polyominoes) with $q^2$ cells that determine the parent exactly. At the powers of two $s \ge q$ tested there are exactly four, the front windows. At other times there are more: some overlap a front window, and others lie at the fronts of the translated sub-configurations described above, e.g. at $\ell_\infty$-distance up to $10$ from the nearest front window for $q=2$, $s=13 = 8 + 4 + 1$. These additional windows do not persist as descendants away from the fronts: at large powers of two, all descendant windows lie at the fronts (Theorem \ref{thm:intrinsic}(ii)). We also verified Proposition \ref{prop:chimera} directly for $2 \le q \le 6$ and $q \le s \le 25$. We conjecture that the front windows, and their counterparts in translated configurations as at $s = 12$ and $s = 20$, are the only such windows for all $q$ and $s$; a proof would require a finer analysis of the parities $N_s(v)$, e.g. by Lucas' theorem.
\end{remark}

The parity rule is often cited as a rule in which every pattern ``reproduces''. In the present framework its behaviour is a fission: the parent's information immediately leaves its original window and travels outward along four fronts, which become causally separated individuals no later than time $q$ and hold exact copies of the parent at times $2^m \ge q$. Whether such a fission counts as \emph{self}-reproduction is a different question, which the genealogy alone cannot answer (Section \ref{sec:discussion}).

\section{The hierarchy of individuals and preferred scales}\label{sec:hierarchy}

The source window of Definition \ref{def:descendant} specifies what is perturbed. Instead of selecting it, we consider all source windows at once. This yields a structure that depends only on the run and the thresholds, in which individuals are nested, and in which the size of the source becomes a further filtration parameter. A selection criterion, autonomy, then marks preferred scales within this structure without defining it.

For the inheritance of descendants by sub-windows we need a uniform version of transmission. For a site $\sigma$, a site $\tau$ not earlier than $\sigma$, and the response map $g$ of Proposition \ref{prop:shannon}, write $\sigma \Rightarrow^*_k \tau$ (\emph{uniform transmission}) if $K(W_\tau\,|\,z) \le k$ and every fibre of $g$ has at most $2^k$ elements. By Proposition \ref{prop:shannon}(i), $\sigma \Rightarrow^*_k \tau$ implies $K(\delta\,|\,\pair{z,v^\delta_\tau}) \le k + c$ for every $\delta$, hence $\sigma \Rightarrow_{k+c} \tau$. Under the parity rule, the front windows are uniformly transmitting at a constant level, since their response maps are bijections (Lemma \ref{lem:front}).

\begin{lemma}[Inheritance by sub-windows]\label{lem:inclusion}
There is a constant $c$ such that the following holds. Let $\sigma = (s, W)$ and $\sigma' = (s, W')$ with $W' \subseteq W$, and let $\tau$ be a site not earlier than $s$. If $\sigma \Rightarrow^*_k \tau$, then $\sigma' \Rightarrow^*_{k + K(W\,|\,W') + c} \tau$.
\end{lemma}
\begin{proof}
Let $z$, $z'$ be the source data of $\sigma$, $\sigma'$, and $y = c_s|_{W \setminus W'}$. The data $z$ are computable from $z'$ and $W$, since the configuration outside $W$ is the restriction of that outside $W'$; hence $K(W_\tau\,|\,z') \le K(W_\tau\,|\,z) + K(W\,|\,W') + c$. The response map of $\sigma'$ is $g'(\delta') = g(\delta' \cup y)$, where $\delta' \cup y$ is the perturbation of $W$ that equals $\delta'$ on $W'$ and $y$ on $W \setminus W'$; so $\delta' \mapsto \delta' \cup y$ maps each fibre of $g'$ injectively into a fibre of $g$, which has at most $2^k$ elements.
\end{proof}

Thus every connected window that receives the perturbations of a window uniformly also receives those of each of its sub-windows, at a level raised by the cost of describing the window from the sub-window; in particular, it contains an inclusion-minimal connected window that does. Descendants of parts can be found inside descendants of wholes. For the fraction-based transmission of Definition \ref{def:transmission}, the analogue can fail: the perturbations of $W'$ form the slice of $\Sigma^W$ in which the rest of the window has its actual content $y$, and a condition on $3/4$ of $\Sigma^W$ says nothing about a particular slice. This is the case of gated transmission, in which whether a part transmits depends on the state of the rest of the window.

\begin{definition}[Canonical descent structure]\label{def:canonical}
For $k \ge 0$, the \emph{descent relation} of a run at level $k$ is the set $\mathcal{T}_k$ of pairs of sites $(\sigma, \tau)$, $\tau$ not earlier than $\sigma$, such that $W_\tau$ is connected and $\sigma \Rightarrow^*_k \tau$. The \emph{individuation structure} of the run is the family of relations $\mathcal{T}_k$, $k \ge 0$, together with the intrinsic genealogies $G_k(\mathcal{D}_{k,\lambda}(\sigma_0, s_1))$ of all sites $\sigma_0$.
\end{definition}

The individuation structure depends only on the run, the rule and the thresholds; no window is chosen. By Lemma \ref{lem:inclusion} it is antitone in the source: enlarging the source window can only remove targets, up to the level shift $K(W\,|\,W') + c$. Individuals come out nested, as parts of individuals with their own descendants, and the size of the source window is a scale parameter alongside $k$ and $\lambda$. By Proposition \ref{prop:invariance}, the structure is invariant up to constant shifts of the thresholds under computable isomorphisms of the intervention structure, since uniform transmission, being a condition on fibre sizes and on tracking, is preserved in the same way as transmission.

Which scales are preferred is a separate question. For a window $W$ and $r \ge 1$, let the \emph{collar} $C_r(W)$ be the set of cells outside $W$ at distance at most $r$ from $W$ in the influence graph.

\begin{definition}[Autonomy]\label{def:autonomy}
A site $\sigma = (s,W)$ is \emph{$(k,r)$-autonomous at time} $s' \ge s$ if there is a connected window $W'$ such that $\sigma \Rightarrow^*_k (s', W')$ and $(s, C_r(W)) \perp_k (s', W')$: the perturbations of the window reach $W'$ in full, and those of its collar, with the window at its actual content, carry at most $k$ bits there.
\end{definition}

Autonomy is an interventional analogue of a boundary or Markov blanket. A scale is preferred in a run if autonomous windows of that size exist over long intervals of time and wide ranges of $k$ and $r$; this is again a persistence notion, and it marks scales rather than defining individuals.

\begin{proposition}[No preferred individuals under the parity rule]\label{prop:noscale}
Under the parity rule, for windows $W$, $W'$ and a lag $t = s' - s$, the fibre sizes of the response map of $(s,W)$ on $(s',W')$, and of that of $(s, C_r(W))$ on $(s',W')$, depend only on $W$, $W'$, $r$ and $t$, and are invariant under translating $W$ and $W'$ together; they do not depend on the run. In particular, up to the constants of Proposition \ref{prop:shannon} and the tracking costs, the collar of $W$ is separated from $W'$ at level $k$ if and only if the rank over $\F_2$ of the linear map $\delta \mapsto F^t(\delta)|_{W'}$, $\delta \in \F_2^{C_r(W)}$, is at most $k$.
\end{proposition}
\begin{proof}
By linearity, the response to a perturbation $\delta$ of a set $U$ is $F^t(\tilde c_s) + F^t(\delta)$, where $\tilde c_s$ is the configuration with $U$ set to $0$; the first term does not depend on $\delta$, so the fibres of the response map on $W'$ are the cosets of the kernel of $\delta \mapsto F^t(\delta)|_{W'}$. This map is determined by the shapes and the lag and commutes with translations. A kernel of dimension $|C_r(W)| - \rho$, where $\rho$ is the rank, gives fibres of size $2^{|C_r(W)| - \rho}$, and Proposition \ref{prop:shannon}(iv) gives $I(\delta : v^\delta\,|\,z) = \rho$ up to $K(W'\,|\,z) + c$ above and $d + c\log(n+2)$ below, for all but a fraction $2^{-d}$ of $\delta$.
\end{proof}

Hence autonomy cannot distinguish the parent's window from any other window of the same shape, in any run of the parity rule: there are no preferred individuals, in accordance with the fact that every pattern is copied in the same way. We computed the rank of the collar map for $q = 4$, $W = Q$, $W' = Q + s e_1$ (the east front window, onto which $W$ transmits uniformly at every $s$) and collars of width $r = 1, 2$:
\begin{center}
\small
\begin{tabular}{lccccccccccccc}
\toprule
$s$ & 4 & 5 & 6 & 7 & 8 & 9 & 10 & 12 & 15 & 16 & 17 & 24 & 32 \\
\midrule
$r = 1$ & 2 & 12 & 8 & 12 & 0 & 12 & 10 & 2 & 12 & 0 & 12 & 0 & 0 \\
$r = 2$ & 8 & 16 & 16 & 12 & 0 & 12 & 16 & 8 & 12 & 0 & 12 & 0 & 0 \\
\bottomrule
\end{tabular}
\end{center}
The collar leaks nothing into the front window at $s \in \{8, 16, 24, 32\}$, in agreement with $F^{2^m} = \sum_e T_{2^m e}$ (and $F^{24} = F^{16}F^{8}$), and leaks up to the full $16$ bits at other times. Autonomy of the front lineages therefore holds only at isolated times, for every window alike.

For autonomy to select a nontrivial family of windows, the rule must contain boundaries. Two naive constructions fail, for instructive reasons. If a membrane is a state of the cells, the interventions include perturbations of the membrane itself, which open it; a random perturbation of a window containing membrane cells almost surely does so. If instead the membrane is built into the law, as cells whose state neither changes nor influences anything, then enclosed regions are simply connected components of the influence graph, and they are selected by the dependence structure alone, without information. A nontrivial example requires a membrane that is a state and is repaired after perturbation, so that the window it encloses is autonomous for most perturbations; this connects autonomy with the closure conditions of autopoiesis \cite{beer2004}.

\section{Illustrations}\label{sec:examples}

\begin{example}[The ship of Theseus]
A ship has its planks replaced one at a time, and the removed planks are stored and later reassembled into a second ship. Model the hull and the stored planks as regions of a deterministic system. Each replacement changes the hull by a bounded amount of information, so the sequence of hull states is a lineage of low cost; the stored planks, carried to their store and reassembled, form another. Before the planks are removed, a perturbation of any part of the ship propagates into both later ships (into the repaired one through the hull, into the reassembled one through the planks), so there is one individual. After removal, perturbations of the stored planks propagate into the reassembled ship and not into the repaired one, and perturbations of the hull and its replacement planks do the reverse. The two lineages are separated, and there are two individuals. In the sense of Definition \ref{def:genealogy}, the original ship has undergone a fission, whose time is well defined at each threshold $k$; the question which of the two ships ``is'' the original has no answer beyond this genealogy, and in particular no answer in terms of the contents of the two ships.
\end{example}

\begin{example}[Reticulation]
Suppose two lineages, separated for a while, come into contact and exchange part of their contents, after which perturbations of each propagate into the other. At the time of contact, two individuals fuse into one; if they later separate again, the genealogy contains a cycle, and the reticulation number is positive. This is the pattern of recombination or of horizontal transfer, as opposed to the purely branching genealogy of Corollary \ref{cor:persistence}.
\end{example}

\section{Adequacy conditions}\label{sec:adequacy}

A definition of individuality is justified by the cases it gets right and by the properties it has. We list the conditions that the present definitions were designed to meet, with the results that establish them, and compare with other approaches.

\begin{enumerate}[label=(A\arabic*)]
\item \emph{Copies.} Two exact copies that cannot affect each other are distinct individuals; copies between which perturbations pass are not certified as distinct. (Lemma \ref{lem:opposite}, the Remark on copies in Section \ref{sec:individuals}, Theorem \ref{thm:fission}.)
\item \emph{Turnover.} A history along which the content changes completely, by few bits per step, while carrying its own perturbations forward, is one individual; information can be lost along it only gradually. (Definition \ref{def:lineage}, Lemma \ref{lem:chain}, the ship of Theseus.)
\item \emph{Fission and fusion.} Divisions and mergers are represented, with a time at each threshold and a graded robustness. (Definition \ref{def:genealogy}, Corollary \ref{cor:persistence}.)
\item \emph{Invariance.} Verdicts do not depend on the universal machine, on computable encodings, or on computable isomorphisms of the intervention structure, up to constant shifts of the thresholds. (Proposition \ref{prop:invariance}; economy is invariant only where the minimal size $m_k$ is locally constant.)
\item \emph{Both ingredients.} The criterion is not a function of the run alone, nor of the run and its dependence structure alone. (Proposition \ref{prop:necessity}.)
\item \emph{Single run.} The criterion applies to one deterministic run with its actual background and history, with no distribution over histories or environments; it quantifies only over perturbations. (Propositions \ref{prop:shannon} and \ref{prop:pointwise}.)
\item \emph{Intrinsic regions.} No region is prescribed: descendant windows are selected intrinsically, and source windows are not chosen but all considered at once, giving a nested structure with the source size as a scale parameter; autonomy marks preferred scales. (Definition \ref{def:descendant}, Theorem \ref{thm:intrinsic}, Definition \ref{def:canonical}, Lemma \ref{lem:inclusion}, Proposition \ref{prop:noscale}.)
\end{enumerate}

Table \ref{tab:comparison} compares the present framework with the ensemble approach of Krakauer et al.\ \cite{krakauer2020}, the interventional information flow of Ay and Polani \cite{aypolani2008} (a measure rather than a theory of individuality), integrated information theory \cite{oizumi2014}, whose exclusion postulate assigns borders to systems, and autopoietic analyses of patterns in the Game of Life \cite{beer2004}. The entries for the other approaches reflect our reading of them and are provisional; they are meant to locate the present proposal, not to evaluate the others on their own aims, several of which differ from ours.

\begin{table}[ht]
\centering
\footnotesize
\setlength{\tabcolsep}{4pt}
\begin{tabular}{lccccc}
\toprule
 & This paper & Krakauer & Ay--Polani & IIT 3.0 & Autopoiesis \\
\midrule
(A1) copies & yes & ? & yes$^a$ & yes & ? \\
(A2) turnover & yes & no$^b$ & n.a. & no$^c$ & yes \\
(A3) fission, fusion & yes & no & n.a. & no$^c$ & ? \\
(A4) invariance & yes$^d$ & partial$^e$ & partial$^e$ & partial$^f$ & ? \\
(A5) both ingredients & yes & no & yes & yes & ? \\
(A6) single run & yes & no & partial$^g$ & partial$^h$ & yes \\
(A7) intrinsic regions & yes$^j$ & partial$^i$ & no & yes & yes \\
\bottomrule
\end{tabular}
\caption{Adequacy conditions (A1)--(A7) for the present framework and other approaches. ``?'': not determined from our reading; ``n.a.'': not applicable, the approach being a measure, not a theory of individuality. Notes: $^a$ used as a separation criterion; $^b$ candidate aggregates are fixed sets of variables, compared from one step to the next; $^c$ the analysis concerns a system at one time; $^d$ up to constants, with the caveat on economy; $^e$ invariant under relabelling of states, but dependent on the choice of variables; $^f$ dependent on the choice of elements and grain; $^g$ the background may be held fixed, but the measure is an average over interventions; $^h$ state-dependent, but defined from the full transition probabilities; $^i$ by optimization over candidate aggregates; $^j$ as a nested structure over all source windows, not as a single choice.}
\label{tab:comparison}
\end{table}

The table is not a ranking. Its point is that conditions (A1)--(A3), which concern identity over time and the distinction of copies, are those that reproduction makes unavoidable, and that the present framework meets them together with (A5) and (A6).

\section{Discussion}\label{sec:discussion}

\emph{What the notion does not decide.} Lineages and genealogies are computed from the run and from interventions on its regions. They describe how information is propagated and separated, i.e. who descends from whom. They do not decide whether an offspring was built by its parent or produced by the dynamical law: under the parity rule, copies arise by a fission that the law performs on any pattern. Deciding this requires interventions on the internal structure of the parent, which is the subject of companion work on self-reproduction.

\emph{Relation to prior work.} Krakauer et al.\ \cite{krakauer2020} define individuality through Shannon information propagated by a candidate aggregate from one time step to the next, conditioned on its environment, which requires an ensemble. Interventional information flow, with background variables either fixed or averaged, was defined by Ay and Polani \cite{aypolani2008}, and its relation to information transfer was analysed by Lizier and Prokopenko \cite{lizier2010}. By Proposition \ref{prop:shannon}, the transmission and separation conditions are, up to additive constants, pointwise versions of Ay--Polani flow with a uniform intervention and a fixed background; we regard them as the interventional core inherited from this work, not as a new measure. Measures that combine information and interventions are also used to delimit systems in integrated information theory and in the effective-information approach to causal emergence; we have not found the two-sided necessity statement of Proposition \ref{prop:necessity} formulated for individuality, but the combination of information and intervention as such is not new. What the present framework adds is: the necessity statement itself; continuity of lineages, which is a condition on individual strings and whose localizing power depends on the complexity of the individual object (Proposition \ref{prop:pointwise}); the tracking condition and the intrinsic selection of descendant regions, including the necessity of connectedness (Proposition \ref{prop:chimera}); the treatment of exact copies; genealogies over long times with fission, fusion and reticulation; the bottleneck filtration and persistence; invariance under computable isomorphisms of the intervention structure (Proposition \ref{prop:invariance}); and theorems for a concrete system (Section \ref{sec:parity}). The use of algorithmic conditional independence for causal questions about individual objects goes back to Janzing and Sch\"olkopf \cite{janzing2010}; the information distance used for links is that of Bennett et al.\ \cite{bennett1998}.

\emph{Invariance.} All conditions are bounds on conditional Kolmogorov complexity, so they are invariant, up to additive constants in the thresholds, under the choice of universal machine and of computable encodings. No metric on space is used: connectedness, steps and light cones are derived from the dependence sets of the rule, so the definitions are invariant under all computable isomorphisms of the intervention structure (Proposition \ref{prop:invariance}). The intervention structure itself, i.e. which sets of states can be set independently, is not determined by information; it is part of what is given, just as an acceptable numbering is given in recursion-theoretic invariance results. Extending invariance to local recodings would require a size measure that does not count cells, e.g. the information $K(v^\delta_\tau\,|\,z)$ carried by the response, in place of $|W|\log_2|\Sigma|$ in the economy condition; we have not checked whether Theorem \ref{thm:intrinsic} survives this change. No window shape is prescribed: descendant windows are selected by tracking, sufficiency, connectedness and economy (Definition \ref{def:descendant}), and the scale is set by the source window and the thresholds $k$ and $\lambda$.

\emph{Why each condition on descendant windows is needed.} Without economy, large simple windows such as the balls of radius $2^n$ in the influence graph qualify; without tracking, scattered sets of cells collected from the whole light cone qualify; without connectedness, chimeras made of parts of separated copies qualify (Proposition \ref{prop:chimera}). Tracking is the only place where Kolmogorov simplicity of the window enters, and it is conditioned on the source data, not on the background alone. It does not penalize a descendant that is carried along by its environment, since the unperturbed configuration is part of the source data.

\emph{Intrinsic source windows.} We do not select source windows; we consider all of them (Section \ref{sec:hierarchy}). The output is therefore a nested structure, not a single list of individuals, which we regard as appropriate: biological individuality is nested, with cells in organisms in colonies. Krakauer et al.\ \cite{krakauer2020} likewise start from candidate aggregates; here the candidates are all windows, and their descendants are derived. Autonomy, an interventional analogue of a Markov blanket, marks preferred scales within the structure. The parity rule has none (Proposition \ref{prop:noscale}), and a rule in which autonomy selects nontrivially needs self-repairing membranes.

\emph{Open problems.}
\begin{enumerate}[label=(\arabic*)]
\item Prove the conjecture of the Remark in Section \ref{sec:parity}, which describes all windows of the parent's size that determine the parent under the parity rule.
\item Determine the intrinsic genealogy $G_k(\mathcal{D}_{k,\lambda}(\sigma_0, s_1))$ of the parity rule, and prove that it has four individuals at times $s \ge q$ (Section \ref{sec:parity}).
\item Construct a rule with self-repairing membranes in which autonomy (Definition \ref{def:autonomy}) provably selects the enclosed regions, and only those, as preferred scales.
\item Find conditions under which the descent relation is inherited by sub-windows for fraction-based transmission, e.g. for most contents of the rest of the window (Lemma \ref{lem:inclusion}).
\item Develop the two-parameter persistence of Section \ref{sec:topology}, and compute it for examples with both fission and fusion.
\item Construct a cellular automaton with provable reticulation, i.e. a genealogy with positive cycle rank.
\item Find a result about individuality in which the algorithmic complexity of the background is essential, beyond what conditioning on a fixed background in the Shannon setting provides; Proposition \ref{prop:pointwise} is of this kind for the parent, not for the environment.
\item Study environments that supply information, e.g. randomly placed building blocks, in which the costs of links, and not only the locality of information, become essential.
\item Relate homotopy classes of chains (Definition \ref{def:homotopy}) to the channels by which information flows in concrete systems, e.g. the four fronts of the parity rule, whose chains from the parent to different copies are not comparable.
\end{enumerate}

\section*{Declaration of generative AI use}

This paper was prepared with substantial assistance from Claude (Anthropic; model Claude Opus 5.5), used through a conversational interface in November 2026. The author proposed tracing descendants through time by information-theoretic means, the analogy with homotopy, the connection with bottleneck distances between programs and proofs, and the replacement of a prescribed window shape by Kolmogorov-simple neighbourhoods; asked for a formal draft; and directed its revision, including the questions of recursive invariance, of the relation to Krakauer et al., and of intrinsic source windows. The AI system formulated the definitions, stated and proved the results, ran the numerical computations reported in the Remark of Section \ref{sec:parity}, suggested the references, and wrote the text.

\bibliographystyle{unsrt}
\bibliography{individuality}

\end{document}